\documentclass{elsart}

\usepackage{natbib}

 \usepackage{graphicx}

\usepackage{amssymb}

\begin{document}

\begin{frontmatter}

 \title{Relativistic Flows at the Hotspots of Radio Galaxies and Quasars?}
 
 \author{Markos Georganopoulos and Demosthenes Kazanas}

 \address{NASA, Goddard Space Flight Center, Code 661, Greenbelt, MD 20771}

\begin{abstract}
We review the broad band  properties of X-ray detected hotspots in radio 
galaxies and quasars. We show that their collective spectral  properties 
can be unified in a framework involving frequency dependent relativistic beaming and varying orientations to the observer's line of sight. The 
simplest dynamic model consistent with this picture is a slowing-down relativistic flow downstream from the hotspot shock, suggesting that the jet
flows remain relativistic to the hotspot distances.

\end{abstract}



\end{frontmatter}

\section{Cygnus A vs. Pictor A}
\label{CygPic}
Pairs of radio emitting jets with lenghts up to several hundred kpc originate
from the central region of radio loud active galaxies. 
In the most powerful of them, the jets terminate in the hotspots, 
compact high brightness regions, where the jet collides with the intergalactic medium (IGM).
 The first hotspots to be detected in X-rays were those of the nearby
powerful radio galaxy Cygnus A \cite{harris94}, whose X-ray flux was found
in agreement with synchrotron self Compton emission in equipartition between
electron and magnetic field energy densities (SSCE). These hotspots 
show no optical emission, suggesting a spectral cutoff at lower frequencies.
While {\it Chandra} observations of Cygnus A confirmed the SSCE picture 
\cite{wilson00}, observations of Pictor A, another nearby powerful 
galaxy \cite{wilson01} showed a very different picture: a.) an one-sided
large scale X-ray jet on the same direction with the known VLBI jet
 \cite{tingay00}; b.) Detection of X-ray emission only from the hotspot 
on the jet side, which was also detected in the optical.
In addition, SSCE models for the Pictor A hotspots underproduce 
the observed X-ray flux, requiring a magnetic field $\sim 14$ times 
below its the equipartition value in order to achieve agreement 
with the observed the X-ray flux.

The one-sided  X-ray jet of Pictor A suggests the potential importance of 
relativistic beaming and orientation as a discriminant of the hotspot
properties of these two sources. An indicator of orientation is the 
ratio $R$ of the core (beamed)   to the extended (isotropic)  radio 
emission. Sources with jets closer to the line of sight are expected to 
have higher  values of R than sources with jets closer to the plane of 
the sky. Cygnus A has $\log R\approx -3.3$ \cite{zirbel95}, while Pictor 
A has  $\log R\approx -1.2 $ \cite{zirbel95}, suggesting that the jets 
of Pictor A are closer to the line of sight than those
of Cygnus A. Another indicator of source orientation is the detection
 of broad
emission lines in the optical-UV spectrum of the core of a source. 
According to the unification scheme for radio  loud active galaxies
(e.g. \cite{urry95}),  broad line 
radio galaxies (BLRG) and  quasars have jets pointing close to the line of sight,
while narrow line radio galaxies (NLRG) have jets closer to the plane of the sky. Cygnus A is a NLRG, and Pictor A is a BLRG,
suggesting again that Pictor A is aligned  closer  to the line of sight.

\section{Collective X-ray--detected hotspot properties; A Unified Picture}
 \label{the rest}

\begin{table}
\caption{Sources with X-ray hotspot detection}
\begin{tabular}{lccccc}
Source	& Type	&	Log R	&          Optical  &	X-ray  & 		SSCE \\
3C 330 & NLRG  &	 -3.5 \cite{saikia94}	&	  NO	&         YES \cite{hardcastle02}, 2 sides & 	   YES \\
Cygnus A  &	NLRG	&	  -3.3\cite{zirbel95}   &	   NO &               YES\cite{harris94,wilson00} 2 sides  &	   YES	\\
3C 295    &	NLRG	 &  -2.7\cite{zirbel95}	&        YES      &        YES \cite{harris00, brunetti01a}, 2 sides 	&   YES	\\
3C 123     &         NLRG  &                -1.9\cite{hardcastle98}      &            NO  &	               YES\cite{hardcastle01}, 1 side    &       YES \\
3C 263          &       Q           &           -1.0\cite{hough89}   &      YES\cite{hardcastle02}, jet side    &  YES\cite{hardcastle02}, Jet side &	   See text \\
3C 351 &   	   Q           &          -1.9\cite{wills86}  &       YES\cite{hardcastle02,brunetti01b}, jet side &  YES\cite{hardcastle02,brunetti01b}, jet side    &    NO \\
Pictor A  &           BLRG      &            -1.2\cite{zirbel95}  &          YES\cite{wilson01}, jet side &     YES\cite{wilson01}, jet side &          NO \\
3C 303  &                Q       &               -0.7\cite{zirbel95}   &        YES\cite{lahteenmaki99}, jet side    &  YES\cite{kataoka03}, jet side   &      NO  \\
3C 390.3    &     BLRG           &         -1.1\cite{zirbel95}    &      YES\cite{lahteenmaki99,prieto97a}, jet side &   YES\cite{prieto97b}, jet side &       NO\\
\end{tabular}
\end{table}

 We examine here the properties of the X-ray detected hotspots in 
relation to the orientation of their  jets relative to the line of sight. 
In table 1 we present the sources  with X-ray hotspot detections, 
their emission line classification, 
core dominance  $R$, their optical and X-ray hotspot properties, 
and the possibility of modeling their spectra with SSCE.
An orientation sequence emerges from these observations: as the jet aligns
closer to the line of sight, measured by an increase in R, the source 
changes from a NLRG 
to a BLRG/Quasar. Sources closer to the plane of the sky show X-ray hotspots
in both lobes (except for the peculiar radio galaxy 3C 123, 
where only one hotspot is detected \cite{hardcastle01}), 
while more aligned sources  show
 X-ray hotspots only on the side of the near jet, as identified
 through VLBI observations.  While in NLRG, SSCE models are  in 
agreement with the observed X-ray flux, in the more aligned 
BLRG and quasars, the X-ray emission from the hotspot is  brighter than the 
SSCE predicted flux, and one has to resort to magnetic fields well 
below equipartition (by a factor of $\sim 10-30$)  to reproduce the 
observed X-ray flux (except for 3C 263, where reproducing the X-ray
 flux requires a magnetic
 field half of the equipartition value\cite{hardcastle02}).
Synchrotron optical emission,  weak or absent for the NLRG hotspots, 
appears at the jet side hotspot 
as the source aligns closer to the line of sight. Finally, radio emission is
present in both hotspots, regardless of orientation, although
 the hotspot in the jet
side of BLRGs and quasars is more powerful and has a flatter spectrum.


\begin{figure}
\centerline{\includegraphics[height=2.4in]{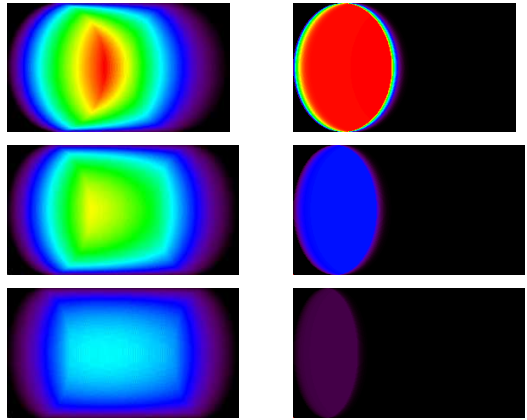}}
\caption{Maps of radio (left) and optical (right) emission under observing angles of 
$40^o$ (top), $50^o$ (middle), and $60^o$ (bottom) from a relativistic 
decelerating flow.}
\end{figure}

This orientation sequence suggests that the synchrotron emission be 
beamed,  with the beaming increasing with frequency. 
The synchrotron optical emission is strongly beamed,
 and in the more aligned objects it is seen in the hotspot of the near jet.
 The radio emission is less beamed, as both radio hotspots are
 seen in all objects.  Increase in alignment leads also to an increase in 
the hotspot X-ray-to-radio ratio, from the SSCE to larger values, 
indicating the presence of an additional component more sensitive to 
orientation effects than the IC component of SSCE. We suggest that 
this frequency dependent beaming results from a relativistic and 
decelerating flow at the hotspots. Such hotspot flow patterns with 
Lorentz factors  up to $\Gamma\sim 3-4$ are routinely seen 
 in relativistic hydrodynamic simulations \cite{aloy99,komissarov96},
and have been used \cite{komissarov96} to explain the fact that, 
in the radio regime,  the near hotspots are more powerful and have a flatter
spectrum than the far ones. 

We show in figure 1 the optical and radio maps
 of a kinematic one-dimensional model
for the hotspot flow. Particle acceleration  takes place at the shock front at
the base of the flow. Electrons are advected downstream loosing energy due
to radiative  losses, while the  bulk flow velocity decelerates
to match the sub-relativistic  advance speed \cite{arshakian00} 
of the hotspots. The most energetic electrons
are located closer to the fast base of the flow and their synchrotron
emission  (mostly optical) is strongly beamed, seen preferentially in the
 near hotspot of the more aligned sources. The radio emission comes  from
the less energetic electrons, further downstream where the flow is slower,
resulting in a wider  beaming pattern, thus detected at a wider range of 
angles. The SSC emission responsible for the X-ray flux (not shown in 
figure 1; work in progress) does not follow the typical SSC/synchrotron
beaming pattern. Because the fast electrons at the base of the flow see the
radio emission of its slower, upstream sections as external radiation, the SSC 
beaming pattern is more focused than the synchrotron one and the 
radio-to-X-ray flux ratio increases as the source aligns to the line of 
sight, while it would remain constant within standard SSC process, 
in support of the slowing-down relativistic flow proposal.

\end{document}